\documentclass[preprint]{elsarticle}
\usepackage{graphicx}
\usepackage{float}
\usepackage[caption = false]{subfig}
\usepackage{amsfonts}
\usepackage{rotating}
\usepackage{amsmath}
\usepackage[compact]{titlesec}
\usepackage{mdwlist}
\usepackage{comment}
\usepackage[labelfont=bf]{caption}

\makeatletter
\renewcommand\@biblabel[1]{}
\makeatother
\makeatletter
\def\ps@pprintTitle{%
  \let\@oddhead\@empty
  \let\@evenhead\@empty
  \let\@oddfoot\@empty
  \let\@evenfoot\@oddfoot
}
\makeatother

\begin{document}
%----------------------------------------------------

\title{Recovering Individual-level Spatial Inference from Aggregated Binary Data}

\author[1]{Nelson B. Walker \corref{cor1}}
\ead{nelsonw@ksu.edu}

\author[1]{Trevor J. Hefley}

\author[2]{Anne E. Ballmann}

\author[2]{Robin E. Russell}

\author[2]{Daniel P. Walsh}

\address[1]{Department of Statistics, Kansas State University, 1116 Mid-Campus Drive North, 
Manhattan, Kansas 66506, U.S.A.}
\address[2]{U.S. Geological Survey, National Wildlife Health Center, 6006 Schroeder Road,
 Madison, WI 53711, U.S.A.}

\cortext[cor1]{Corresponding author}

\begin{abstract}
Binary regression models are commonly used in disciplines such as epidemiology and ecology to determine how spatial covariates influence individuals. In many studies, binary data are shared in a spatially aggregated form to protect privacy. For example, rather than reporting the location and result for each individual that was tested for a disease, researchers may report that a disease was detected or not detected within geopolitical units. Often, the spatial aggregation process obscures the values of response variables, spatial covariates, and locations of each individual, which makes recovering individual-level inference difficult. We show that applying a series of transformations, including a change of support, to a bivariate point process model allows researchers to recover individual-level inference for spatial covariates from spatially aggregated binary data. The series of transformations preserves the convenient interpretation of desirable binary regression models that are commonly applied to individual-level data. Using a simulation experiment, we compare the performance of our proposed method under varying types of spatial aggregation against the performance of standard approaches using the original individual-level data. We illustrate our method by modeling individual-level probability of infection using a data set that has been aggregated to protect an at-risk and endangered species of bats. Our simulation experiment and data illustration demonstrate the utility of the proposed method when access to original non-aggregated data is impractical or prohibited.

\end{abstract}

%  Please place your key words in alphabetical order, separated
%  by semicolons, with the first letter of the first word capitalized,
%  and a period at the end of the list.
%

%  As usual, the \maketitle command creates the title and author/affiliations
%  display 

\maketitle

%  If you are using the referee option, a new page, numbered page 1, will
%  start after the summary and keywords.  The page numbers thus count the
%  number of pages of your manuscript in the preferred submission style.
%  Remember, ``Normally, regular papers exceeding 25 pages and Reader Reaction 
%  papers exceeding 12 pages in (the preferred style) will be returned to 
%  the authors without review. The page limit includes acknowledgements, 
%  references, and appendices, but not tables and figures. The page count does 
%  not include the title page and abstract. A maximum of six (6) tables or 
%  figures combined is often required.''

%  You may now place the substance of your manuscript here.  Please use
%  the \section, \subsection, etc commands as described in the user guide.
%  Please use \label and \ref commands to cross-reference sections, equations,
%  tables, figures, etc.
%
%  Please DO NOT attempt to reformat the style of equation numbering!
%  For that matter, please do not attempt to redefine anything!

\section{Introduction}
\label{s:intro}

%\linenumbers

Spatially referenced binary data are among the most common types of data that enable inference about spatial covariates. Scientists and policy makers are often interested in understanding how spatial covariates influence the probability of a binary outcome, such as whether a plant or animal tests positive or negative for a disease. Sometimes spatial binary data are aggregated to protect privacy. For example, wild plants and animals are protected by law (e.g., threatened or endangered species under the U.S. Endangered Species Act (ESA) of 1973). As a result, spatially referenced binary data involving protected plants and animals may be reported in aggregate to reduce the potential for  human contact (e.g. tourism, vandalism, and theft). The aggregation process can make individual-level inference difficult to obtain for spatial covariates because the original values of the binary responses, locations, and spatial covariates cannot be recovered.

\begin{figure}
\centerline{%
\includegraphics[width=150mm,trim={0cm 0cm 0cm 0cm},clip]{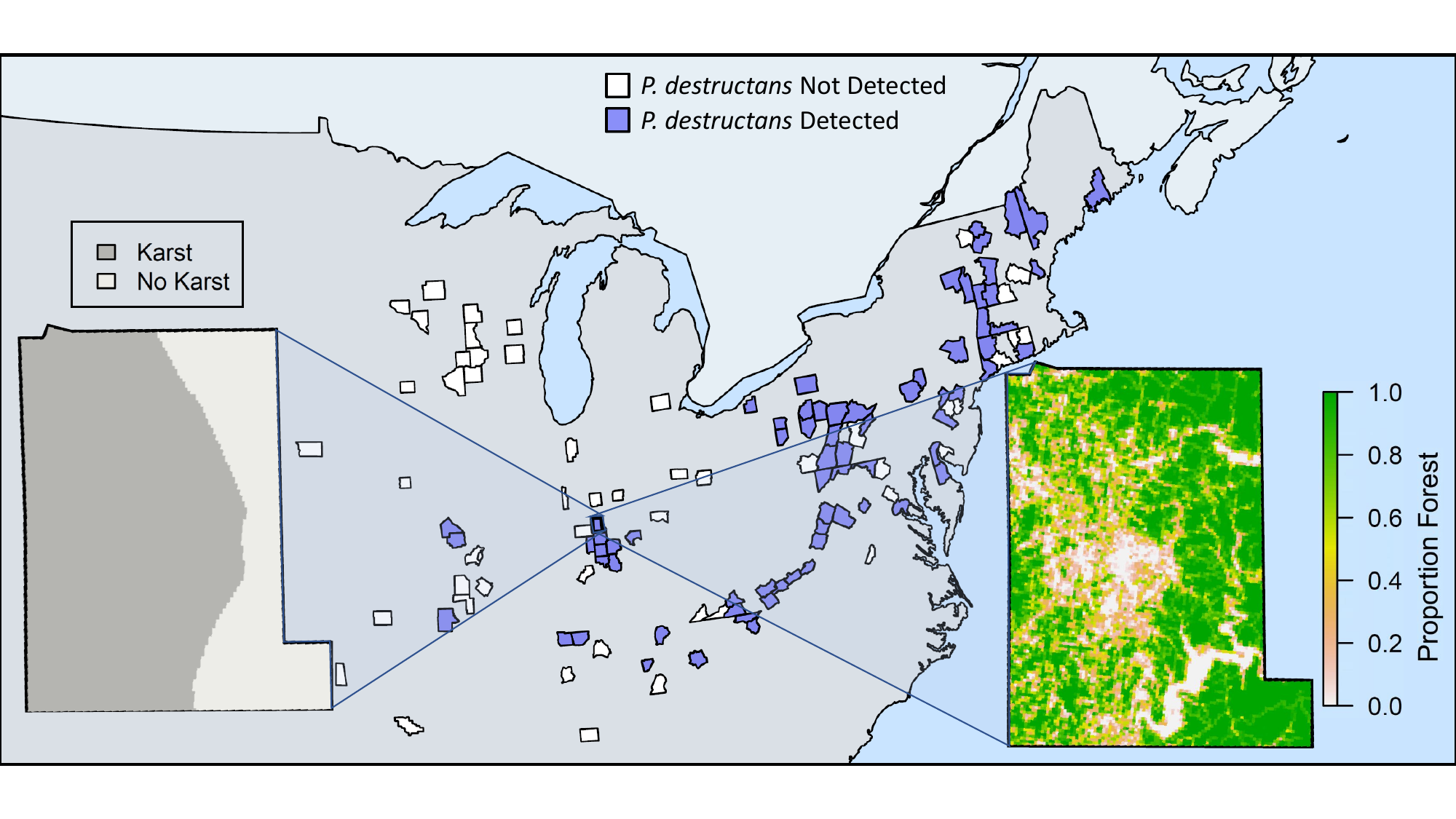}}
\caption{The motivating data set shows which counties contained bats that were individually tested for \textit{P. destructans}, the causative agent of white-nose syndrome, within the northeastern United States from 2008-2012. The counties that contained at least one bat that tested positive for \textit{P. destructans} are shown in purple fill while counties with no positive bats are shown in white fill. The covariates `proportion of land classified as forest' (inset right) and `presence of karst' (inset left) from Monroe county, Indiana, USA (outlined in bold black). Karst is a type of landscape characterized by caves and sinkholes that can provide  habitat to cave-hibernating bats. Spatially referenced wildlife data are often accessible to researchers in aggregated form to reduce the potential for human contact. When binary data within a county are aggregated into an indicator that denotes whether the county contained at least one sampled bat that tested positive, individual-level spatial covariates and inference cannot be obtained. %Researchers commonly resort to fitting regression models to the areal-level data and may attempt to interpret the areal-level inferences as if they were individual-level inferences on spatial covariates.
}
\label{fig:location-error-panel}
\end{figure}

An example where spatial binary data are aggregated is a disease surveillance study for white-nose syndrome (WNS), which is caused by the fungal pathogen \textit{P. destructans}. In a disease surveillance study, binary observations are collected on individual bats found within geopolitical areas (counties). However, the observations are aggregated to the county-level when making them accessible to researchers and the public in accordance with federal law and to protect the wildlife (see \textbf{Figure 1}). The map in \textbf{Figure \ref{fig:location-error-panel}} indicates which counties in the northeastern United States contained individual bats that were tested and which counties had at least one diagnosed case of WNS from 2008-2012. When the individual test results are aggregated as shown in \textbf{Figure \ref{fig:location-error-panel}}, it can be difficult to recover the original individual-level inference for spatial covariates because the original values of the binary response, location, and spatial covariates for each observation are unknown. For these types of data, researchers commonly resort to fitting regression models to the aggregated data and may interpret the areal-level inference about spatial covariates as if it was obtained from a model that was fit to individual-level data, which is a well-documented ecological fallacy (Piantadosi, Byar, and Green, 1988; Gotway and Young, 2002).

% new paragraph%
%Spatially aggregated binary data may also arise from spatial group-based testing. Group testing is the act of taking individual samples and compiling them into a group or pool that is then tested for the presence of a disease or chemical. When the group is defined by boundaries of a geopolitical unit, the result is a spatial group test. Group testing is advantageous because it requires only a single test for the entire group rather than testing each individual sample in the group. Methods that match spatial group testing and aggregated binary data with an appropriate model are needed to recover individual-level inference on spatial covariates. Recently, there has been interest in using the aggregated binary data that arise from group testing to make individual-level inference regarding covariates. However, the individual-level covariate values must be accessible or obtained at the time of data collection (e.g., Vansteelandt, Goetghebeur, and Verstraeten, 2004; McMahan et al., 2017; Joyner et al., 2019). If the individual covariate values were unavailable at the time of sample collection, spatial group testing and privacy protection result in equivalent forms of aggregated binary data. 

\begin{figure}
\centerline{%
\includegraphics[width=190mm,trim={0cm 1.5cm 0cm 1cm},clip]{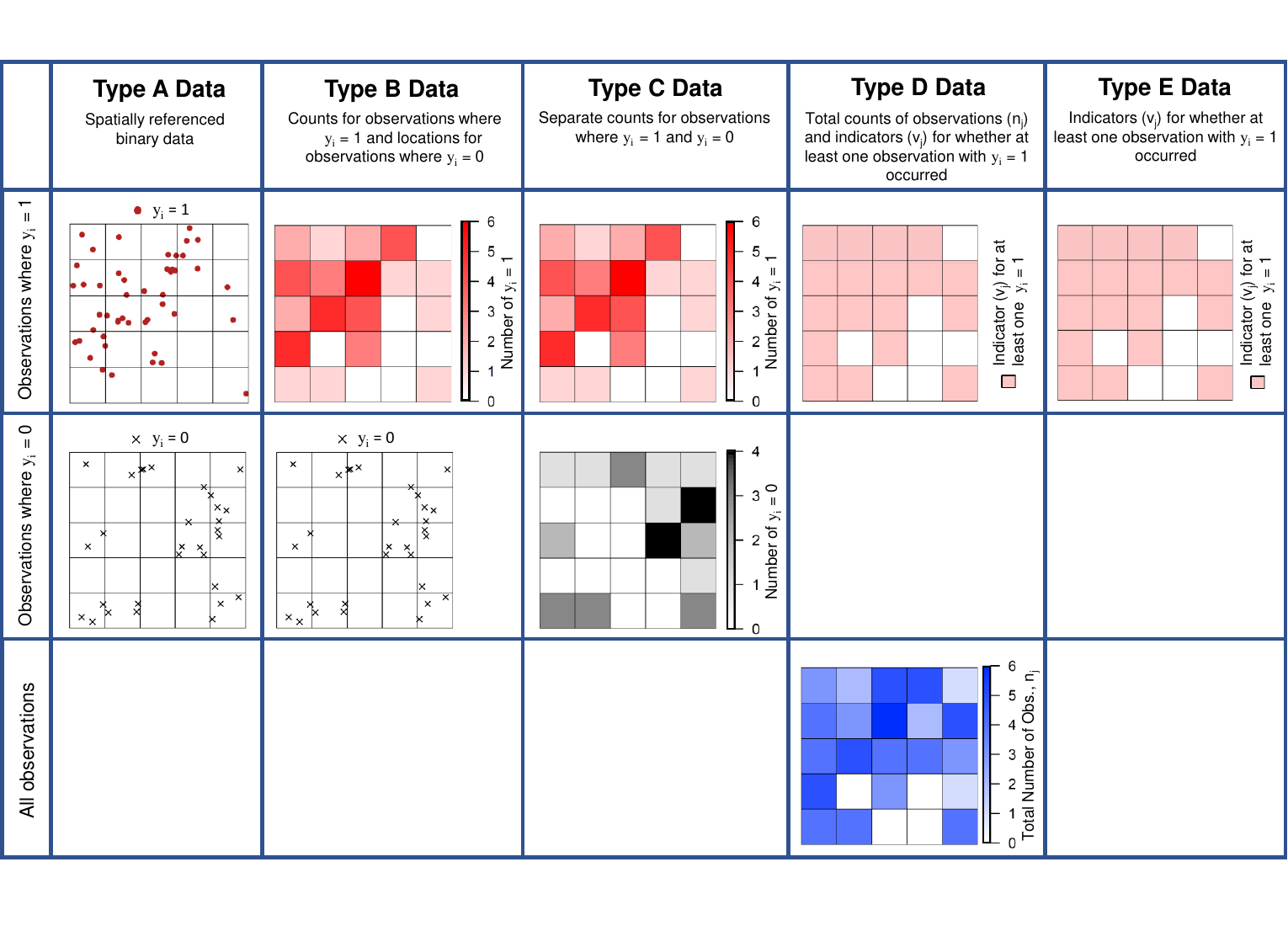}}
\caption{Graphical representations of the types of aggregation for spatially referenced binary data found in \textbf{Table 1}. The data set shown under Type A is progressively aggregated across sub-regions, starting from the exact locations of all observations (Type A data) and ending with binary indicators (Type E data). We define $y_i$ as the binary mark associated with  the $i^\mathrm{th}$ spatially referenced observation. For the $j^\mathrm{th}$ subregion, we  define $n_j$ as the total number of observations contained therein. We also define $v_j$ as a binary indicator that at least one observation with $y_i=1$ occurred in the $j^\mathrm{th}$ subregion. 
}
\label{agg_types}
\end{figure}

Univariate point process-based methods have traditionally formed the backbone of efforts to make individual-level inference on spatially aggregated data (e.g., Bradley et al., 2016; Hefley et al., 2017; Taylor, Andrade-Pacheco, and Sturrock, 2018; Gelfand and Shirota, 2019). Perhaps less common, bivariate point process models enable individual-level inference on spatially aggregated data where the non-aggregated data consist of binary marks at specific locations (Diggle et al., 2010a; Chang et al., 2015; Wang et al., 2017; Johnson, Diggle, and Giorgi, 2019; Walker, Hefley, and Walsh, 2020). For binary data, these methods are capable of recovering individual-level inference on spatial covariates under varying types of spatial aggregation (see \textbf{Table \ref{aggregation_data}} and \textbf{Figure \ref{agg_types}}). For example, when the individual-level binary data are aggregated over areal units into separate counts of the number of observations with a specific binary mark, the methods by Wang et al. (2017), Johnson et al. (2019), and Walker et al. (2020) can be used to recover individual-level inference for spatial covariates (see \textbf{Table \ref{aggregation_data}}, Type C). When at least some of the binary data are aggregated into counts (e.g., number of observations with a mark of zero) and the rest of the data are not aggregated, the methods from Diggle et al. (2010a), Chang et al. (2015), and Walker et al. (2020) can be used to recover individual-level inference for spatial covariates (see \textbf{Table \ref{aggregation_data}}, Type B). 

\begin{sidewaystable}
\caption{Different types of aggregation or privacy protection for spatially referenced binary data, along with their relative information content, and references that successfully recover individual-level inference for each type of data. See \textbf{Figure 2} for visualizations of the aggregation types.
} 
\label{aggregation_data}%l@{\extracolsep{\fill}}
{\begin{tabular*}{\columnwidth}{@{}c@{\extracolsep{\fill}}l@{\extracolsep{\fill}}c@{\extracolsep{\fill}}l@{\extracolsep{\fill}}c@{}} 
\hline    
			&Aggregation/																	&Information						& Example	References That							\\
Type		&Privacy Protection																&Content							& Enable Individual-level Inference				\\ \hline \hline
%A			&None																			&Complete							&Waller and Gotway (2004)							\\ 
A			&None																				&Complete							&Diggle and Giorgi (2019)		 					\\ 
B			&Some non-aggregated data w/ subregion counts of ones or zeros	&High								&Diggle et al. (2010a)									\\  
			&																						&										&Chang et al. (2015)									\\ 
			&																						&										&Walker et al. (2020)									\\ 
C			&Subregion counts of ones and zeros  									&High								&Wang et al. (2017)									\\ 
			&								  														&										&Johnson et al. (2019)									\\ 
			&								  														&										&Walker et al. (2020)									\\ 
D			&Total subregion counts and subregion indicator of ones			&Medium							&\textbf{No methods currently exist}			\\ 
			&								  														&										&										 						\\	
E			&Subregion indicator of ones or zeros				 					&Low									&\textbf{No methods currently exist}			\\ \hline 
%Synthetic																&High								&(To Generate): Wang and Reiter, 2012; Paiva et al., 2014; Quick et al., 2015; Yu et al., 2017	
\end{tabular*}}
\bigskip
\end{sidewaystable}

Aside from Type B and C data, we have identified two additional types of aggregated data that appear in practice.  First, when the data are aggregated into counts of the total number of observations in areal units and also aggregated into binary indicators that denote whether at least one observation in the areal unit had a mark of one, the existing methods are insufficient to recover individual-level inference on spatial covariates (see \textbf{Table 1} and \textbf{Figure \ref{agg_types}}, Type D). Likewise, to the best of our knowledge, no methods exist to recover individual-level inference on spatial covariates when the aggregated data consist only of the binary indicators over areal units (see \textbf{Table 1} and \textbf{Figure \ref{agg_types}}, Type E). This is unfortunate because, presumably, data categorized as Type D or E are more likely to be accessible when compared to  data of Type B or C. We hypothesize that Type D and Type E data would be more accessible because Type D and E are a degraded form of Types A-C data and offer a higher degree of privacy protection. Thus, Type D and E aggregated data are an untapped data source for individual-level inference. For example, the disease surveillance example from \textbf{Figure \ref{fig:location-error-panel}} may be classified as Type E data. 

The contribution of this paper is to enable individual-level inference for spatial covariates from Type D and E aggregated binary data. We accomplish this by transforming the bivariate inhomogeneous Poisson point process (BIPPP) regression model and applying several distributional results. Importantly, and following Walker et al. (2020), the proposed methods preserve the interpretation of commonly used binary regression methods (e.g., logistic and probit regression). Thus the proposed methods are easy to interpret and are widely applicable to aggregated binary data.

The remainder of this paper proceeds as follows: In the Methods Section, we review regression models for binary data, including the BIPPP. We then present several distributional results for the transformed BIPPP that may be used to recover individual-level spatial inference under various types of aggregation. In the Simulation Experiment Section, we evaluate and compare the proposed models to traditional approaches for the analysis of spatial binary data (e.g., logistic regression) using a simulation study. In the Application Section, we apply our proposed regression models to a data example from wildlife disease surveillance where the aggregated data result in a binary indicator for each geopolitical unit. Finally, in the Discussion Section, we identify potential modifications and model comparisons that practitioners may consider.

\section{Methods}
\label{s:ipp}

\subsection{Binary Regression}

Binary regression is arguably one of the most popular types of regression models and can be written as 

\begin{align}
\label{eq:binary_reg}
y_i \sim &
\mathrm{Bernoulli}(p_i), \\
\label{eq:g_link}
g(p_i)=&
\beta_0 + \mathbf{{x}}_i^{\prime}\boldsymbol{\beta},  
\end{align}

\noindent where $y_i$ is the $i^{\mathrm{th}}$ binary response from $\mathbf{y} \equiv (y_1,y_2,\dotsc,y_n)^\prime$, $n$ is the number of observations, $p_i$ is the probability that $y_i=1$, and $g(\cdot)$ is an appropriate link function (e.g., logit or probit). Additionally, $\beta_0$ is an intercept, $\mathbf{{x}}_i\equiv (x_1, x_2,\dotsc,x_q)^\prime$ is a vector of $q$ covariates, and $\boldsymbol{\beta} \equiv (\beta_1, \beta_2,\dotsc,\beta_q)^\prime$ is a vector of $q$ regression coefficients. Regression models like (\ref{eq:binary_reg}-\ref{eq:g_link}) are often used to model spatial binary data (e.g., Gelfand and Schliep, 2018; Diggle and Giorgi,  2019). In the case that (\ref{eq:g_link}) includes spatial covariates $\mathbf{x(\mathbf{s})}$, then $p_i$ becomes a spatially varying function such that 

\begin{align}
\label{eq:g_spatial_link}
g(p(\mathbf{s}))=\beta_0 + \mathbf{x}(\mathbf{s})^{\prime}\boldsymbol{\beta}, 
\end{align}

\noindent where $\mathbf{s} \equiv (s_1,s_2)^\prime$ is a coordinate vector within the study area  $ \mathcal{S}$ (i.e., $\mathbf{s} \subseteq \mathcal{S}$). In what follows, we specify $g(\cdot)$ using the logit link function, however, as with any binary regression model, the choice is flexible.

A similar spatial binary regression model to (\ref{eq:binary_reg}) and (\ref{eq:g_spatial_link}) that incorporates the locations of $n$ observations in a study area $\mathcal{S}\subset \mathbb{R}^2$, is the bivariate point process (Gelfand and Schliep, 2018). Perhaps the most common type of point process used for binary data is the bivariate inhomogeneous Poisson point process (BIPPP; Gelfand and Schliep, 2018). The BIPPP is a joint distribution composed of a Poisson probability mass function that generates $n$, a location density that generates the coordinates of each observation, $\mathbf{u}_i$, and the Bernoulli probability mass function from (\ref{eq:binary_reg}) that generates binary outcomes, $\mathbf{y}$, called marks (Gelfand and Schliep, 2018).  The BIPPP can be written as:

\begin{align}
\label{eq:biv_IPP}
f(n,\mathbf{{u}}_1,\mathbf{{u}}_2,\dotsc,\mathbf{{u}}_{n},\mathbf{y}|\lambda,p) = &
  \frac{e^{-(\int_\mathcal{S}{\lambda(\mathbf{{s}})d\mathbf{{s}}})}{(\int_\mathcal{S}{\lambda(\mathbf{{s}})d\mathbf{{s}}})}^{n}}{n!} \nonumber
  \times \\
& \prod\limits^{n}_{i=1}\frac{\lambda(\mathbf{{u}}_{i})
 }{\int_\mathcal{S}{\lambda(\mathbf{{s}})%p(\bmath{\mathrm{s}})
 d\mathbf{{s}}}} 
  \hspace{5px} p(\mathbf{{u}}_{i})^{y_i}(1-p(\mathbf{{u}}_{i}))^{1-y_i}\hspace{1pt},
\end{align}

\noindent where $\lambda(\cdot)$ is a spatially varying thinned intensity function that captures both the distribution of bats and the sampling process (Gelfand and Shirota, 2019). The function $p(\cdot)$ is identical to (\ref{eq:g_spatial_link}) and may be viewed as a classification function because it relates a binary mark to each of $n$ locations. For example, in our motivating data set, the binary marks represent test results for individual bats that tested positive (i.e., $y_i=1$) or negative ($y_i=0$) for \textit{P. destructans}, the causative agent of WNS. We note that the BIPPP offers no obvious advantage for spatial binary data over the model formed from (\ref{eq:binary_reg}) and (\ref{eq:g_spatial_link}) unless the binary observations are spatially aggregated, the locations of the observations are obscured by location error (e.g., Walker et al., 2020), or the observations are collected via preferential sampling (e.g., Diggle, Menezes, and Su, 2010b).

In many applications, researchers often specify $\lambda(\cdot)$ using

\begin{equation}
\label{eq:loglink}
\mathrm{log}(\lambda(\mathbf{s}))=\alpha_0 + \mathbf{z}(\mathbf{s})'\boldsymbol{\alpha}\hspace{2pt},
\end{equation}

\noindent where $\alpha_0$ is an intercept, $\mathbf{z}(\mathbf{s}) \equiv (z(\mathbf{s})_1, z(\mathbf{s})_2,\dotsc,z(\mathbf{s})_r)^\prime$ is a vector of $r$ spatial covariates, and $\boldsymbol{\alpha} \equiv (\alpha_1, \alpha_2,\dotsc,\alpha_r)^\prime$ is a vector of $r$ regression coefficients (Gelfand and Schliep, 2018). Some situations may require an alternative, and potentially more flexible, specification in (\ref{eq:loglink}). For example, a Gaussian process could be added to (\ref{eq:loglink}) by way of a spatial random effect (Gelfand and Schliep, 2018). We focus on a log-linear specification for $\lambda(\cdot)$ because the specification is reasonable for our motivating data set and because we can more easily discover parameter identifiability issues.

\subsection{Change of Support and Distributional Results} 
  
While the distributions from (\ref{eq:binary_reg}) and (\ref{eq:biv_IPP}) are appropriate for spatially referenced binary data, they are inadequate when the spatial binary data are aggregated (see \textbf{Table 1}). In what follows, we outline several transformations of the BIPPP that result in distributions that match the distributional attributes of aggregated spatial binary data of Types C, D, and E (see \textbf{Table 1} and \textbf{Figure 2}).

The transformation of a spatial process from continuous to areal support is called a change of support (COS). To implement a COS, the study area $\mathcal{S}$ is partitioned into $J$ non-overlapping subregions, $\mathcal{A}_1, \mathcal{A}_2,\dotsc,\mathcal{A}_J$, such that $\mathcal{S}=\cup^J_{j=1} \mathcal{A}_j$. The partition is determined by how the data were aggregated. 
For example, our motivating data set reported the county that each bat was sampled from in the northeastern United States  (see \textbf{Figure \ref{fig:location-error-panel}}). 
Thus, $\mathcal{S}$ is defined by the combined area of the counties that contained sampled bats and the partition is defined by the boundaries of the counties which contained the bats. 

If we know the number of observations with a mark of one ($n_{1j}$) and a mark of zero ($n_{0j}$) contained within the $j^{\mathrm{th}}$ subregion (see \textbf{Table 1} and \textbf{Figure 2}, Type C data), a result of applying the COS to the BIPPP is $n_{1j}$ and $n_{0j}$ are Poisson random variables distributed as follows (Gelfand and Schliep, 2018): 

\begin{align}
\label{eq:COS_pos}
n_{1j} \sim &
\mathrm{Pois}(\int_{\mathcal{A}_j}{\lambda(\mathbf{s})p(\mathbf{s}})d\mathbf{s}),\\
\label{eq:COS_neg}
n_{0j} \sim &
\mathrm{Pois}(\int_{\mathcal{A}_j}{\lambda(\mathbf{s})(1-p(\mathbf{s}))}d\mathbf{s}).
\end{align}

\noindent The joint distribution of $n_{1j}$ and $n_{0j}$ is an appropriate density for binary data that have been aggregated into counts and results in a regression model that recovers individual-level inference on  spatial covariates. Effectively, this models two point patterns, with intensities $\lambda(\mathbf{s})p(\mathbf{s})$ and $\lambda(\mathbf{s})(1-p(\mathbf{s}))$, for presence and absence of a mark. Wang et al. (2017) and Walker et al. (2020) both used this type of binary regression model to make individual-level inference from aggregated binary data using spatial covariates. 
 Similar to (\ref{eq:COS_pos}-\ref{eq:COS_neg}), the number of observations in the $j^{\mathrm{th}}$ subregion, $n_j=n_{1j}+n_{0j}$, is also a Poisson random variable (Cressie and Wikle, 2011, p. 207), 

\begin{align}
\label{eq:COS_all}
n_j \sim &
\mathrm{Pois}(\int_{\mathcal{A}_j}{\lambda(\mathbf{s}})d\mathbf{s}).
\end{align}

\subsubsection{Proposed Change-of-Support based Methods}

In some cases, we may have access to $n_j$ (e.g., the total number of individuals tested within each county) and a binary indicator $v_j = \mathrm{I}(n_{1j}>0)$ for each subregion (see \textbf{Table 1} and \textbf{Figure 2}, Type D data). In our motivating data set, $v_j=1$ indicates that the $j^{\mathrm{th}}$ county contains at least one sampled bat that tested positive for the pathogen, and $v_j=0$ indicates that all of the sampled bats tested negative in the county. 
Conditioning $v_j$ on $n_j$, we obtain the following density: 

\begin{align}
\label{eq:latent2.bivar.ipp}
v_j|n_j\sim &
\textrm{Bern}(1-(1-\tilde{p}_j)^{n_j}), 
\end{align}
\noindent where
\begin{align}
\label{eq:define.p.tilde}
\tilde{p}_j = &
\frac{\int_{\mathcal{A}_j}{\lambda(\mathbf{s})p(\mathbf{s}})d\mathbf{s}}{\int_{\mathcal{A}_j}{\lambda(\mathbf{s}})d\mathbf{s}}.
\end{align}

\noindent The conditional distribution of $v_j$ given $n_j$ is an appropriate density for binary data that have been aggregated into Type D data. The joint density of (\ref{eq:COS_all}) and (\ref{eq:latent2.bivar.ipp}) can also be used to construct a regression model for Type D aggregated binary data. Models based on (\ref{eq:latent2.bivar.ipp}) or the joint distribution of (\ref{eq:COS_all}) and (\ref{eq:latent2.bivar.ipp}) are a novel development because both can recover individual-level inference on spatial covariates from Type D aggregated data (see \textbf{Table 1}). 

Under the form of aggregation in Type E data, we may assume only $v_j$ is given for each subregion (see \textbf{Table 1} and \textbf{Figure 2}). The data generated by the indicator function follow a Bernoulli distribution and is given as follows:

\begin{align}
\label{eq:latent1.bivar.ipp}
v_j\sim \textrm{Bern}(1-e^{-\int_{A_j}\lambda(\mathbf{s})p(\mathbf{s})d\mathbf{s}}).
\end{align}
A model for Type E data based on (\ref{eq:latent1.bivar.ipp}) is also a novel development, as the model is capable of recovering individual-level inference on spatial covariates from Type E aggregated data.

\subsection{Parameter Identifiability}

The distributions presented in Section 2.2 form the basis for regression models that recover individual-level spatial inference from various types of aggregated binary data (see \textbf{Table 1} and \textbf{Figure 2}). Like all binary regression models and point process models, the proposed transformed BIPPP models may have parameter identifiability issues (e.g., complete separation; Hefley and Hooten, 2015) when sample size is small or the data contain little information (e.g., a very large number of zeros).

\subsection{Model Implementation}
\label{model.implementation}
We use the Nelder-Mead algorithm in the program R to numerically minimize the negative log-likelihoods for the densities introduced in this paper and simultaneously estimate all parameters (R Core Team, 2020). Evaluating the negative log-likelihood functions requires approximating the integrals contained therein. We approximate the integrals using simple quadrature for ease of implementation (e.g., $\int_{\mathcal{A}_j} \lambda(\mathbf{s})d\mathbf{s} \approx \sum_{k=1}^K |W|*\lambda(\mathbf{s}_k)$, where $\lambda(\mathbf{s}_k)$ is the value of $\lambda(\mathbf{s})$ at the $k^\mathrm{th}$ quadrature point and $|W|$ is the area of a grid cell that is both a subset of $\mathcal{A}_j$ and approximated by a quadrature point). 
%Other integral approximations (e.g., Gaussian quadrature) may be more efficient but more complex to program. 
For all model parameters, we approximate variances by inverting the Hessian matrix and then construct Wald-type confidence intervals (CIs).

\section{Simulation Experiment}
\label{s:simulation}

We conducted a simulation experiment to compare the performance of our proposed models, using different types of aggregated binary data, to traditional models for non-aggregated binary data (e.g., logistic regression). We simulated data using a unit square study area, $\mathcal{S}=[0,1] \times [0,1]$, that was divided into $400$ regular grid cells (subregions), such that $\mathcal{S}=\cup^{400}_{j=1} \mathcal{A}_j $ and $|\mathcal{A}_j|=\frac{1}{400} $. We generated spatial covariates, $x(\mathbf{{s}})$ and $z(\mathbf{{s}})$, and simulated the locations and binary marks of observations from a BIPPP where the intensity function was log$(\lambda(\mathbf{s}))=\alpha_0+\alpha_1z(\mathbf{s}) $ and the classification function was logit$(p(\mathbf{s}))=\beta_0+\beta_1x(\mathbf{s})$. 
We focused on and compared estimates of $\beta_1$ among five models because $\beta_1$ is highly affected by aggregation and inference on the slope parameters of the classification function are likely to be the focus of many applied studies (Walker et al., 2020). %Additionally, studies that use our proposed methods are likely to focus on inference for parameters associated with spatial covariates in the classification function.
%We focused on and compared estimates of $\beta_1$ among five models since this parameter is the focus of most applied studies. 
We accomplished the comparison of estimates of $\beta_1$ by assessing bias, coverage probabilities (CPs), and relative efficiency for estimates of $\beta_1$ among the following five scenarios: 
\begin{enumerate*}
\item[1.] A traditional logistic regression model from (\ref{eq:binary_reg}) and (\ref{eq:g_spatial_link}) fit to non-aggregated data (see \textbf{Table 1} and \textbf{Figure 2}, Type A);
\item[2.] A joint model for $n_{1j}$ and $n_{0j}$ that is specified by combining the distributions in (\ref{eq:COS_pos}) and (\ref{eq:COS_neg}; see \textbf{Table 1} and \textbf{Figure 2}, Type C);
\item[3.] A joint model for $v_j$ and $n_j$ that is specified by combining the distributions in (\ref{eq:COS_all}) and (\ref{eq:latent2.bivar.ipp}; see \textbf{Table 1} and \textbf{Figure 2}, Type D); 
\item[4.] The conditional model for $v_j$ given $n_j$ from (\ref{eq:latent2.bivar.ipp}; see \textbf{Table 1} and \textbf{Figure 2}, Type D);  
\item[5.] The Bernoulli model for $v_j$ from (\ref{eq:latent1.bivar.ipp}; see \textbf{Table 1} and \textbf{Figure 2}, Type E).

\end{enumerate*}

We simulated 1000 data sets from four different settings using a combination of two factors: covariate equivalence ($x(\mathbf{{s}})=z(\mathbf{{s}})$ vs.  $x(\mathbf{\mathrm{s}})\ne z(\mathbf{{s}})$); and average sample size (small vs. large). Thus our simulation experiment uses a total of 4,000 simulated data sets and realizations of $z(\mathbf{s})$ and $x(\mathbf{s})$. Each simulated data set was aggregated to fit each data type in scenarios 2-5. We drew each spatial covariate realization from a low-rank Gaussian process (Higdon, 2002) on a $200 \times 200$ grid with knots at every fourth grid cell to reduce computation time. We chose parameter values of $\alpha_1=1$, and $\beta_1=1$ for all settings. We chose values for $\alpha_0$ and $\beta_0$ for each setting such that the average sample size per subregion was either 10 or 50 (small vs. large) and the proportion of subregions that contained a binary mark of one was approximately constant across all settings. The values of $\alpha_0$ and $\beta_0$ in settings 1-4 were $7.800, 9.410, 7.820, 9.405$ and $-5.500, -7.070, -4.750, -6.350$, respectively.

We fit the model in scenario one (i.e., traditional logistic regression) using the \texttt{glm} function in R to obtain the maximum likelihood estimates (MLEs) of $\beta_0$ and $\beta_1$. We fit the models in scenarios two through five as described in Section 2.4. For each model and setting, we calculated and compared the CPs from the 95$\%$ Wald-type CIs for $\beta_1$. We also constructed box plots comparing the distribution of $\hat{\beta}_1$ obtained from the 1000 data sets for each scenario and setting. 
We calculated the standard deviation of the empirical distribution of the 1000 estimates of $\beta_1$ in each scenario. We then calculated the relative efficiency of $\hat{\beta}_1$ for scenarios two through five by dividing the standard deviation of the distribution of $\hat{\beta}_1$ for the respective scenario by that of scenario one.
Lastly, we calculated the mean squared predictive error (MSPE) in the estimated intensity and probability surfaces for each of the models in scenarios two through five. However, we only calculated the MSPE for the estimated probability surface for the model in scenario one.

When binary data are generated according to a BIPPP and then spatially aggregated, we expect to obtain unbiased estimates in scenarios two, three, four, and five. 
%However, in settings where $\mathrm{x}(\mathbf{s})= \mathrm{z}(\mathbf{s})$, we expect the parameters in the model from scenario five to be very weakly identifiable. 
Of the proposed models based on the distributional results presented in Sections 2.1-2, we expect that the model for scenario two will have the highest relative efficiency 
%and the fewest issues with identifiability 
among all settings covered by the experiment, followed by the models from scenarios three, four, and five. We expect the MSPE of the estimated intensity and probability surfaces to be smallest for the model in scenario two, followed by three, four, and five.
 We provide annotated R code capable of reproducing the simulation experiment in the \texttt{simulation.R} file in the supporting information. %Plots showing results from all settings in the simulation experiment are provided in Web Appendix A. 

\subsection{Simulation Results}

\renewcommand{\thesubfigure}{\Alph{subfigure}}
\begin{figure}
%\centerline{
\centering
\hspace*{\fill}
\subfloat{\includegraphics[width = 2.5in]{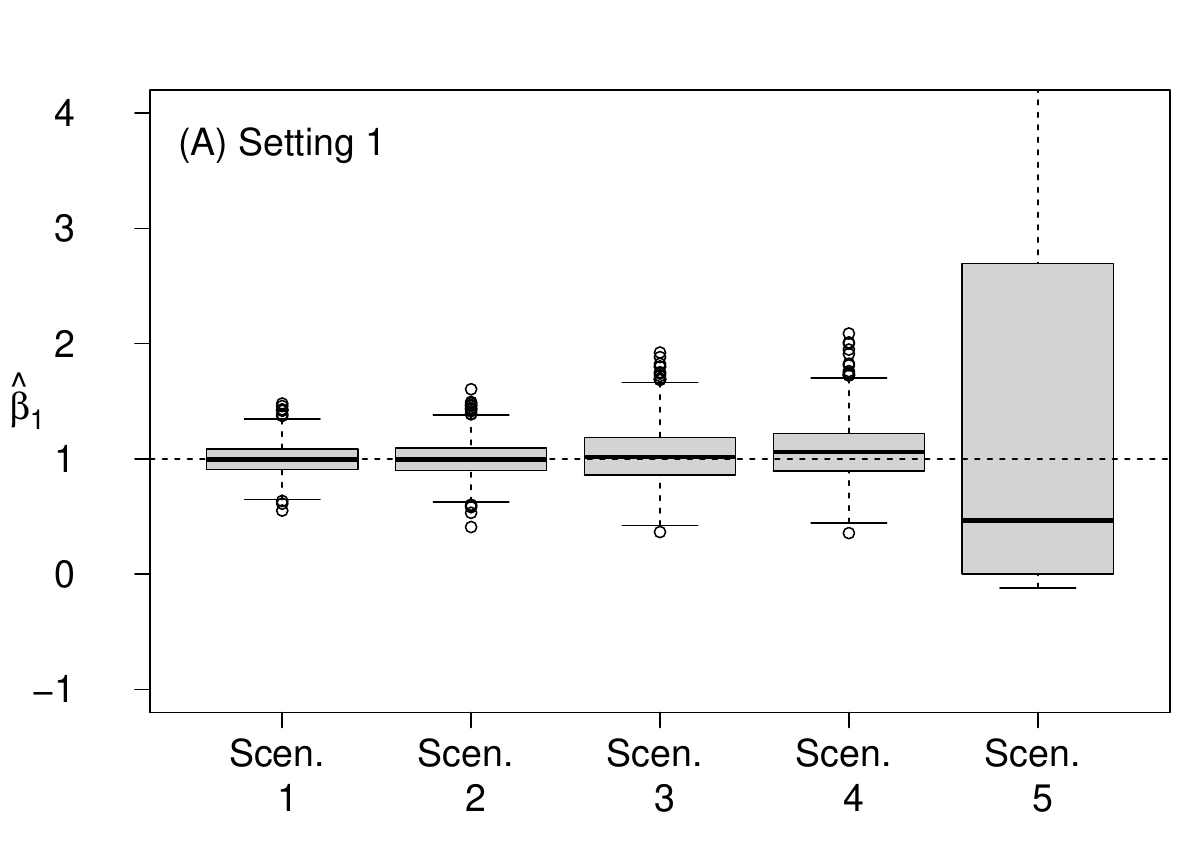}}
\hspace*{\fill} 
\subfloat{\includegraphics[width = 2.5in]{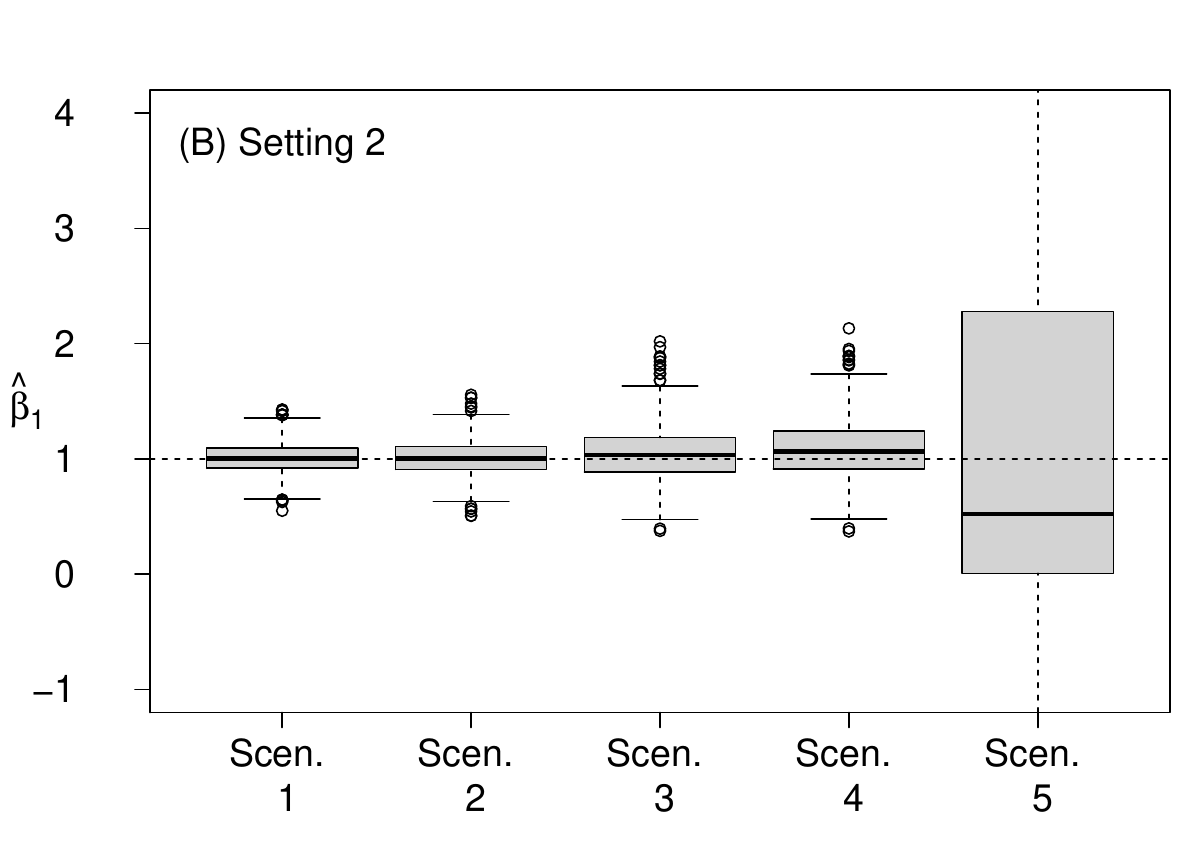}} \hspace*{\fill}
\\ \hspace*{\fill}
\subfloat{\includegraphics[width = 2.5in]{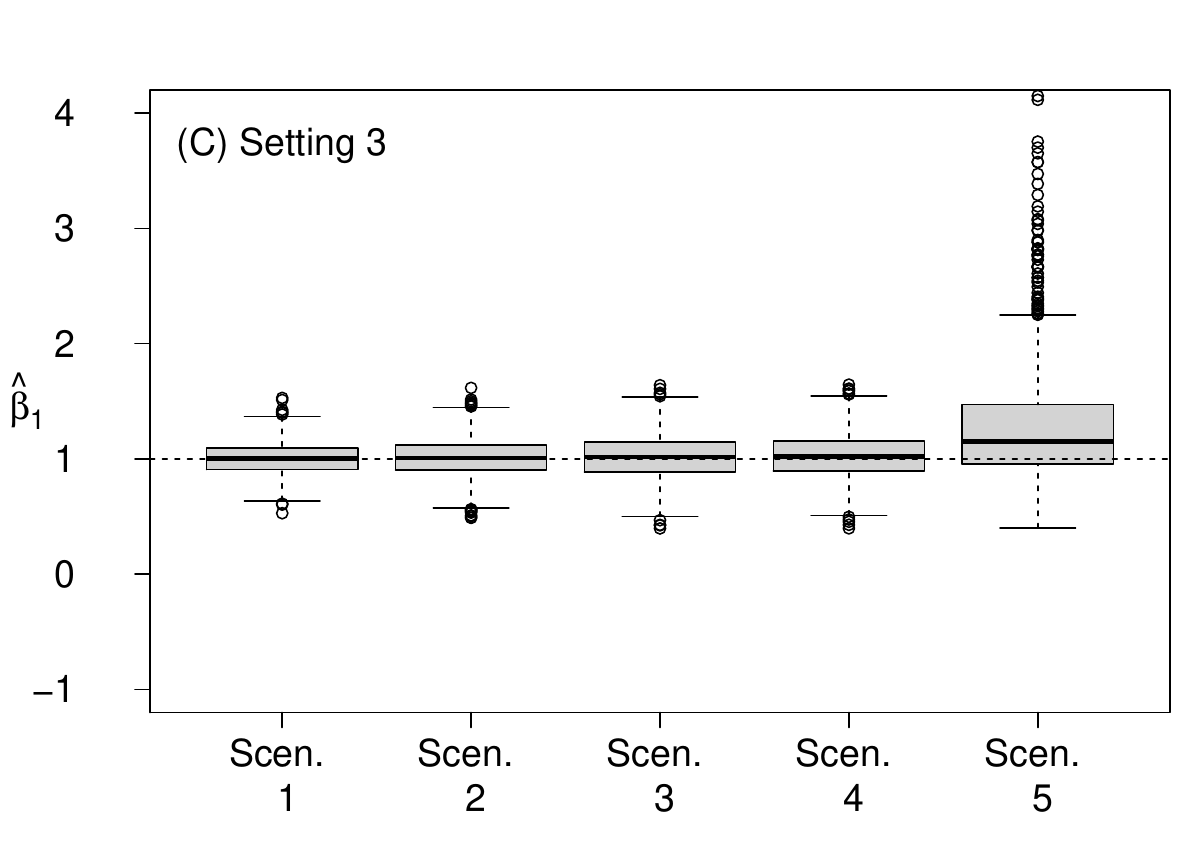}}
\hspace*{\fill}
\subfloat{\includegraphics[width = 2.5in]{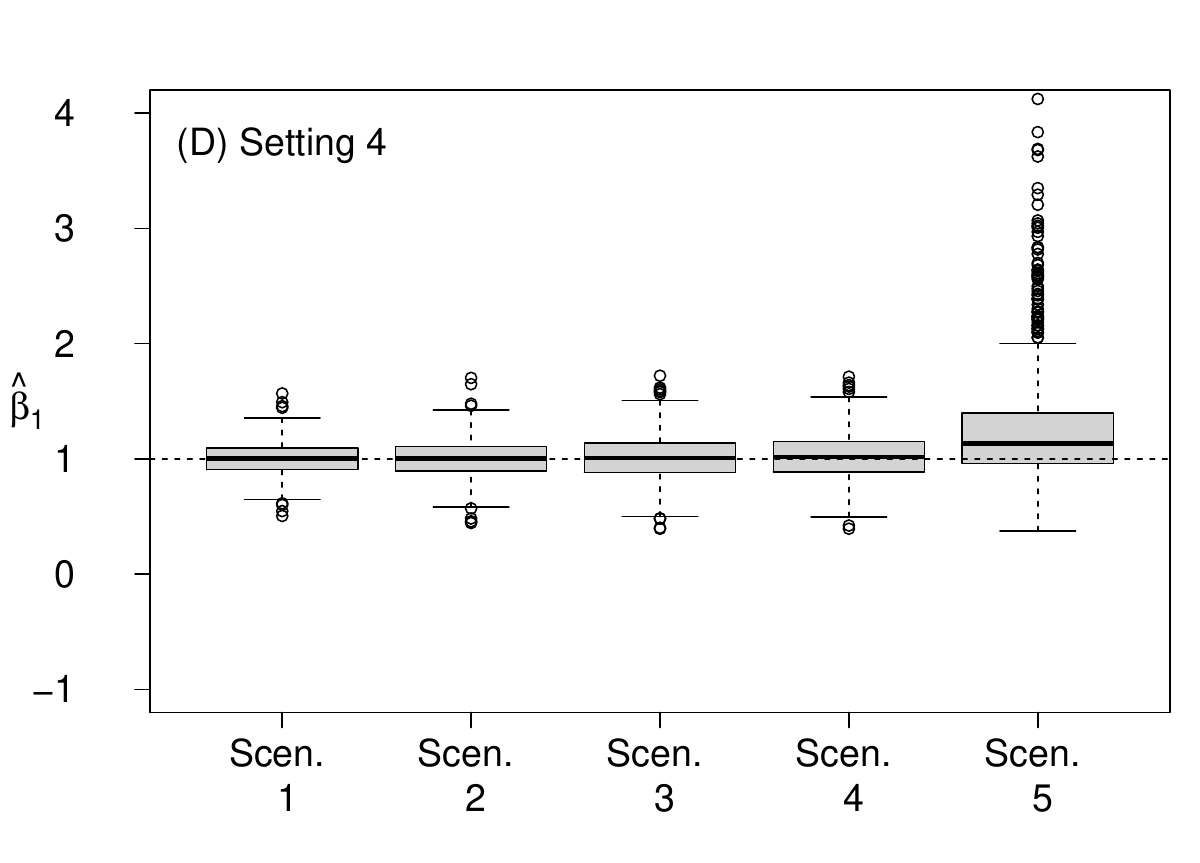}}
\hspace*{\fill}
 %}
%\includegraphics[width=180mm,trim={0 8cm 0cm 0},clip]{scenarios.pdf}}
%\includegraphics[width=100mm,trim={0 1cm 0cm 0},clip]{Violin_comparison_COS_logit.pdf}}
%\includegraphics[width=100mm,trim={0 1cm 0cm 0},clip]{setting2.pdf}}
%% to include a figure, or to leave a blank space
\caption{Panels (A) and (B) show box plots of results from small and large average sample size simulation experiment settings where $x(\mathbf{s})=z(\mathbf{s})$. Panels (C) and (D) show small and large sample size simulation experiments where $x(\mathbf{s}) \ne z(\mathbf{s})$. We show maximum likelihood estimates of $\beta_1$ obtained using five different models (each under a different data aggregation scenario), which included: \textbf{Scen. 1)} logistic regression with no data aggregation (Type A data); \textbf{Scen. 2)} a joint model for $n_{1j}$ and $n_{0j}$ where binary data were aggregated into counts for each subregion (Type C data); \textbf{Scen. 3)} a joint model for $v_{j}$ and $n_{j}$ using data aggregated into a count and indicator variable for each subregion (Type D data); \textbf{Scen. 4)} a conditional model for $v_{j}$ given $n_{j}$ using data aggregated into a count and indicator variable for each subregion (Type D data); \textbf{Scen. 5)} a Bernoulli model for $v_{j}$ using data aggregated into an indicator variable for each subregion (Type E data). Each of the four panels used 1,000 simulated data sets, and each panel shows the true value of $\beta_1=1$ (dotted line). The distribution of $\hat{\beta}_1$ from scenario five (Bernoulli model) was such that some estimates fell outside the upper bounds of the plots. Each box plot shows (from bottom to top) the lower bound of 1.5 times the inter-quartile range, the 25$^\mathrm{th}$ percentile, the median, the 75$^\mathrm{th}$ percentile, and the upper bound of 1.5 times the inter-quartile range.  See \textbf{Table 2} for a summary of all settings. 
}
\label{fig:COS simulation}
\end{figure}
%MAYBE TALK ABOUT ALL THE NA's FROM MODEL 8 AND 10

 In our simulation experiment, we crossed two factors (average sample size per subregion and covariate equivalence) with two levels each. With our choices of $\alpha_0$, the average numbers of observations within each grid cell were about 10.2 and 50.1 for small and large sample settings, respectively. With our choices of $\beta_0$ for each setting, we maintained a proportion of approximately 0.11 of grid-cells that contained a binary mark of one (see \textbf{Table \ref{sim.results}}). 
 
As expected, under the model and data in scenario one (traditional logistic regression with no data aggregation), the MLEs for $\beta_1$ appear to be unbiased for all settings and had CPs between $0.945$ and $0.951$. Under the model and data in scenario two (joint distribution of $n_{1j}$ and $n_{0j}$) the MLEs for $\beta_1$ appear to be unbiased for all settings in the simulation study (see \textbf{Figure 3} for graphical comparisons of estimates and the web-based appendix for additional plots and summaries). The CPs for $\hat{\beta}_1$, in scenario two, were between $0.94$ and $0.961$ for all settings. Additionally, the relative efficiency of $\hat{\beta}_1$, obtained from scenario two, ranged from about 1.1 (settings 1, 2) to about 1.2 (setting 3). The CPs obtained for scenarios one and two, and efficiencies for scenario two, are available in \textbf{Table 2}.

Under the model and data in scenario three (joint distribution of $v_{j}$ and $n_{j}$) the MLEs for $\beta_1$ appear to be unbiased for all settings in the simulation study (see \textbf{Figure 3}). 
The CPs for $\hat{\beta}_1$, in scenario three, were between $0.946$ and $0.964$ for all settings. Additionally, the relative efficiency of $\hat{\beta}_1$, obtained from scenario three, ranged from about 1.4 (setting 4) to about 1.8 (setting 2). The CPs and efficiencies obtained for scenario three are available in \textbf{Table 2}.

Under the model and data in scenario four (conditional distribution of $v_{j}$ given $n_{j}$) the MLEs for $\beta_1$ appear to be unbiased for all settings in the simulation study (see \textbf{Figure 3}). 
The CPs for $\hat{\beta}_1$, in scenario four, were between $0.919$ and $0.943$ for all settings. Additionally, the relative efficiency of $\hat{\beta}_1$, obtained from scenario four, ranged from about 1.4 (setting 4) to about 1.9 (setting 2). Finally, under the model and data in scenario five (Bernoulli distribution of $v_{j}$), the MLEs for $\beta_1$ were weakly identifiable with efficiencies of $\hat{\beta}_1$ ranging from about 13.1 (setting 4) to over 18,000 (setting 3) and CPs between $0.819$ and $0.956$. The CPs and efficiencies obtained for scenarios four and five are available in \textbf{Table 2}. 

As expected, the MSPE of the estimated probability surfaces was smallest for the model in scenario one, followed by two, three, four, and five across all settings. In general, the MSPE of the estimated intensity surfaces were smallest for the model in scenario two, followed by three, four, and five. Plots showing the distributions of the MSPE for the estimated intensity and probability surfaces among each of the scenarios for all settings are given in the web-based appendix.

\begin{sidewaystable}
%\begin{table}
\caption{Results from our simulation experiment using two sample sizes (small vs. large) and two levels of covariate equivalence ($x(\mathbf{{s}})=z(\mathbf{{s}})$ vs. $x(\mathbf{{s}})\ne z(\mathbf{{s}})$). For each setting, we report the average number of observations within each grid cell ($\bar{n}_j$), the average number per grid cell that had a mark of one ($\bar{n}_{1j}$) and zero ($\bar{n}_{0j}$), and the average proportion of grid cells that contained an observation with a mark of one $(\bar{v}=\frac{1}{1000}\frac{1}{400}\sum_{sim=1}^{1000}\sum_{j=1}^{400}\mathrm{I}(n_{1j}>0))$ from 1,000 simulated data sets. We show the relative efficiency (Eff.) for estimating $\beta_1$ and the 95\% CI coverage probability (CP) for each of the proposed models (using appropriate types of aggregated data). We also report the 95\% CI CP for logistic regression using the exact locations of observations. We calculate the relative efficiency for each proposed model as the ratio of the standard deviation of the empirical distribution of $\beta_1$ from the respective proposed model against that of logistic regression.
%We calculate efficiency as the mean ratio of the standard errors of $\alpha_1$ from the bivariate COS (under location error) and logistic regression (when exact locations are known).  
} 
\label{sim.analysis}
{\begin{tabular*}{\columnwidth}{@{}c@{\extracolsep{\fill}}c@{\extracolsep{\fill}}c@{\extracolsep{\fill}}c@{\extracolsep{\fill}}c@{\extracolsep{\fill}}c@{\extracolsep{\fill}}c@{\extracolsep{\fill}}c@{\extracolsep{\fill}}c@{\extracolsep{\fill}}c@{\extracolsep{\fill}}c@{\extracolsep{\fill}}c@{\extracolsep{\fill}}c@{\extracolsep{\fill}}c@{\extracolsep{\fill}}c@{\extracolsep{\fill}}c@{}}
\hline     
			&Covariate										&				&						&							&							&														&	CP			&  CP			&  CP			&  CP			&  CP			&	Eff.		&	Eff.		&	Eff.		&	Eff.			\\				
			&Equivalence							 		& Sample	&						&							&							& 														&Scen.		&Scen.		& Scen.		&	Scen.		&Scen.		&	Scen.		&Scen.		& Scen.		&	Scen.			\\  
Setting	&($x(\mathbf{s}) = z(\mathbf{s})$) &	Size 		&$\bar{n}_j$		&$\bar{n}_{1j}$		&	$\bar{n}_{0j}$	&$\bar{v}$										&1				&2				&3				&4				&5				&2				&3				&4				&5				\\ \hline
1			& Yes											& Small		& 10.1				&	0.18					&	9.89					& 0.11												& 0.951 		& 0.960		& 0.964 		&	0.919		& 0.875		&	1.12		&	1.75		& 1.81		&	1,606			\\ 	
2			& Yes											& Large		& 50.3				&	0.19					&	50.1					& 0.11												& 0.950		& 0.961		& 0.958 		&	0.932		& 0.819		&	1.12		&	1.78		& 1.86		&	248.4			\\
3			& No												& Small		& 10.2				&	0.14					&	10.1					& 0.11												& 0.945 		& 0.940 		& 0.946 		& 	0.922		& 0.956		&	1.22		&	1.41		&	1.43		&	18,516			\\ 
4			& No												& Large		& 49.9				&	0.14					&	49.8					& 0.11												& 0.951 		& 0.955 		& 0.959 		&	0.943		& 0.940		&	1.15		&	1.37		&	1.39		&	13.13		\\

 \hline  
\end{tabular*}}
\bigskip
\label{sim.results}
%\end{table}
\end{sidewaystable}

\section{Application} 

\subsection{Disease Risk Factor Analysis}
\label{s:cwd application}

The distributional results outlined in the Methods Section are useful for disease risk factor analyses when data have been spatially aggregated. Using the transformed distributions enables researchers to recover individual-level inference about how spatial covariates influence the probability of infection. We illustrate our proposed methods using disease surveillance data collected to understand and manage the spread of white-nose syndrome (WNS). As previously mentioned, WNS is a fungal disease caused by the pathogen 
\textit{P. destructans} 
%\textit{Pseudogymnoascus destructans} 
 that threatens several North American species of bats (Ingersoll, Sewall, and Amelon, 2016). The earliest documentation of the disease in North America was in 2006 based on photographic evidence from Howes Cave, near Albany, New York (Blehert et al., 2009; Frick et al., 2010; Hefley et al., 2020). The pathogen, \textit{P. destructans}, has since spread throughout the eastern and midwestern United States resulting in high mortality rates among several species of cave-hibernating bats. Surveillance for \textit{P. destructans} in the United States began in 2007 using a combination of passive and active surveillance methods. During 2007--2012, samples were obtained from individual bats associated with morbidity or mortality investigations occurring year-round at underground hibernacula or on the above-ground landscape. An individual sample consisted of a bat carcass, biopsies of wing skin, or tape lifts of fungal growth on the muzzle. A small number of individual samples were also obtained from target species (including \textit{Myotis} spp., \textit{Perimyotis subflavus}, and \textit{Eptesicus fuscus}) that were admitted to rehabilitation facilities or state diagnostic laboratories for rabies testing from approximately December to May. A positive or negative diagnosis of WNS in individual bats was determined by observing characteristic histopathologic lesions in skin tissues using light microscopy (Meteyer et al. 2009). A diagnosis of `suspect WNS' was assigned to individuals with clinical signs suggestive of the disease that had ambiguous skin histopathology or that had the causative agent (\textit{P. destructans}) detected by fungal culture, fungal tape lift, or polymerase chain reaction in the absence of available or definitive histopathology and regardless of observed clinical signs (Lorch et al. 2010). We treated `suspect WNS' diagnoses as positive cases for our analysis.

We illustrate our modeling approach using a subset of the WNS surveillance data collected during 2008--2012 that included individual samples of little brown bats (\textit{Myotis lucifugus}), big brown bats (\textit{Eptesicus fuscus}), northern long-eared bats (\textit{Myotis septentrionalis}), and tri-colored bats (\textit{Perimyotis subflavus}). 
This resulted in %at least fifty individual samples for each species and 
a total of 428 samples with 226 positive or suspected positive cases of WNS (Ballmann et al., 2021). 
%In our analysis, we chose to exclude samples from 2006-2007 because diagnostic case criteria for WNS were not established until 2008. 
As a result of the data collection process, the study area $\mathcal{S}$ was defined as the 120 counties that contained at least one bat that was tested for WNS between 2008 and 2012. The resulting study area collectively covered approximately 195,000 km$^2$. We note that this number reflects the sum of the areas of the included counties rather than the area of the northeastern United States. %The locations of all bats in the study area follow a point process. The locations of sampled bats then represent a thinned point process (Fithian and Hastie, 2013). 
To comply with the Endangered Species Act and protect the bats and their environment, the locations of the tested bats were recorded as the respective county centroids and thus suffered from bounded location error (\textit{sensu} Walker et al., 2020). As bounded location error is equivalent to aggregation in this instance, the original data are Type C and require an appropriate model (i.e. the joint model for $n_{1j}$ and $n_{0j}$ from  (\ref{eq:COS_pos}-\ref{eq:COS_neg})) to obtain bias corrected individual-level inference. As Type C data can be further aggregated to become Type D and E, the WNS data are well-positioned to demonstrate our proposed models.

We were interested in two spatial covariates when we evaluated our proposed models. The first spatial covariate was `presence of karst' (karst), a type of landscape characterized by cave formation. Therefore, the presence of karst in any particular area serves as a plausible surrogate covariate for the presence or absence of caves where bats might congregate (Medellin, Wiederholt, and Lopez-Hoffman, 2017). The second spatial covariate was `proportion of land classified as forest' (forest) and was calculated from the 2011 National Land Cover Database by determining what proportion of land within each $300\mathrm{m}\times300$m grid cell in the study area was composed of any kind of forest (MLRC, 2011). The forest covariate is notable because the proportion of the immediate vicinity that is covered in forest may be an ecologically relevant predictor for the presence of WNS (Jachowski et al., 2014).

We fit each of four regression models that enable individual-level spatial inference from aggregated binary data (i.e., the joint model for $n_{1j}$ and $n_{0j}$ from (\ref{eq:COS_pos}-\ref{eq:COS_neg}); the joint model for $v_j$ and $n_j$ from  (\ref{eq:COS_all}) and (\ref{eq:latent2.bivar.ipp}); the conditional model for $v_j$ given $n_j$ from (\ref{eq:latent2.bivar.ipp}); and the Bernoulli model for $v_j$ from (\ref{eq:latent1.bivar.ipp})) to the WNS data set under the types of aggregation introduced in \textbf{Table 1} (Types C, D, and E). We incorporated the spatial covariate `presence of karst' in the thinned intensity function, $\lambda(\mathbf{s})$, of the proposed transformed models and we included `proportion of land classified as forest' (forest) as the spatial covariate in $p(\mathbf{s})$ in the transformed models.

%We fit each of four regression models that enable individual-level spatial inference from aggregated binary data (i.e., the joint model for $n_{1j}$ and $n_{0j}$ from (\ref{eq:COS_pos}-\ref{eq:COS_neg}); the joint model for $v_j$ and $n_j$ from  (\ref{eq:COS_all}) and (\ref{eq:latent2.bivar.ipp}); the conditional model for $v_j$ given $n_j$ from (\ref{eq:latent2.bivar.ipp}); and the Bernoulli model for $v_j$ from (\ref{eq:latent1.bivar.ipp})) to the WNS data set under the types of aggregation introduced in \textbf{Table 1} (Types C, D, and E). We incorporated the spatial covariate `presence of karst' in the thinned intensity function, $\lambda(\mathbf{s})$, of the proposed transformed models. This was appropriate because karst terrain is an indicator of cave formation in the eastern U.S., making karst a plausible surrogate covariate for the presence or absence of caves where bats might congregate. We included `proportion of land classified as forest' (forest) as the spatial covariate in $p(\mathbf{s})$ in the transformed models. The forest covariate was calculated from the 2011 National Land Cover Database by determining what proportion of land within each $300\mathrm{m}\times300$m grid cell in the study area was composed of any kind of forest (MLRC, 2011). The forest covariate is notable because the proportion of an immediate vicinity that is covered in forest may be an ecologically relevant predictor for the presence of WNS (Jachowski et al., 2014). 

We also fit three logistic regression models to the Type E aggregated WNS data, consisting of indicator variables (see \textbf{Table 1}, Type E). These three models represent the approach some researchers resort to when attempting to make individual-level inference from aggregated data. The first model that was fit to Type E data used the value of the forest covariate from the centroid of each county (Areal County Centroid), while the second model used the average of the forest covariate for each county (Areal County Average). The third logistic regression model that was fit to Type E data used the average of the forest covariate across areas in each respective county where karst landscape was present (Areal $\%$ Forest in Karst). %These models that were fit to areal-level data represent the approach some researchers resort to when attempting to make individual-level inference from aggregated data. %Lastly, we fit a traditional logistic regression model to the non-aggregated WNS data set using the forest covariate values for each observation.

We fit the regression models that enable individual-level spatial inference from aggregated binary data as outlined in Section \ref{model.implementation} using the program R. We used the \texttt{glm} function in the program R to fit the specified logistic regression models (R Core Team, 2020). Numerically optimizing the likelihood functions for the proposed regression models each required approximately one and a half hours on a standard desktop computer. We compare MLEs and Wald-type 95\% CIs among the proposed regression models and we provide the MLEs and Wald-type 95$\%$ CIs for the three logistic regression models fit to Type E data as a reference. We provide annotated R code capable of reproducing the disease risk factor analysis in the \texttt{wns.R} file in the supporting information and in Ballmann et al. (2021).

\subsection{Results}
\label{s:results}

\begin{figure}
\centerline{%
%\includegraphics[scale=0.5]{Covariate_Panel.eps}
%\fbox{\includegraphics[scale=.5]{CovariatePanel.eps}}} 
%\includegraphics[width=170mm,trim={0 0cm 2cm 0},clip]{LocationError3panelcovariateplot.pdf}}
%\includegraphics[width=170mm,trim={10cm 40cm 40cm 2cm},clip]{test.pdf}}
%\includegraphics[width=170mm,trim={4.8cm 4cm 6cm 1.5cm},clip]{composite1_out.pdf}}
\includegraphics[width=150mm,trim={.5cm 0.5cm 0cm 0cm},clip]{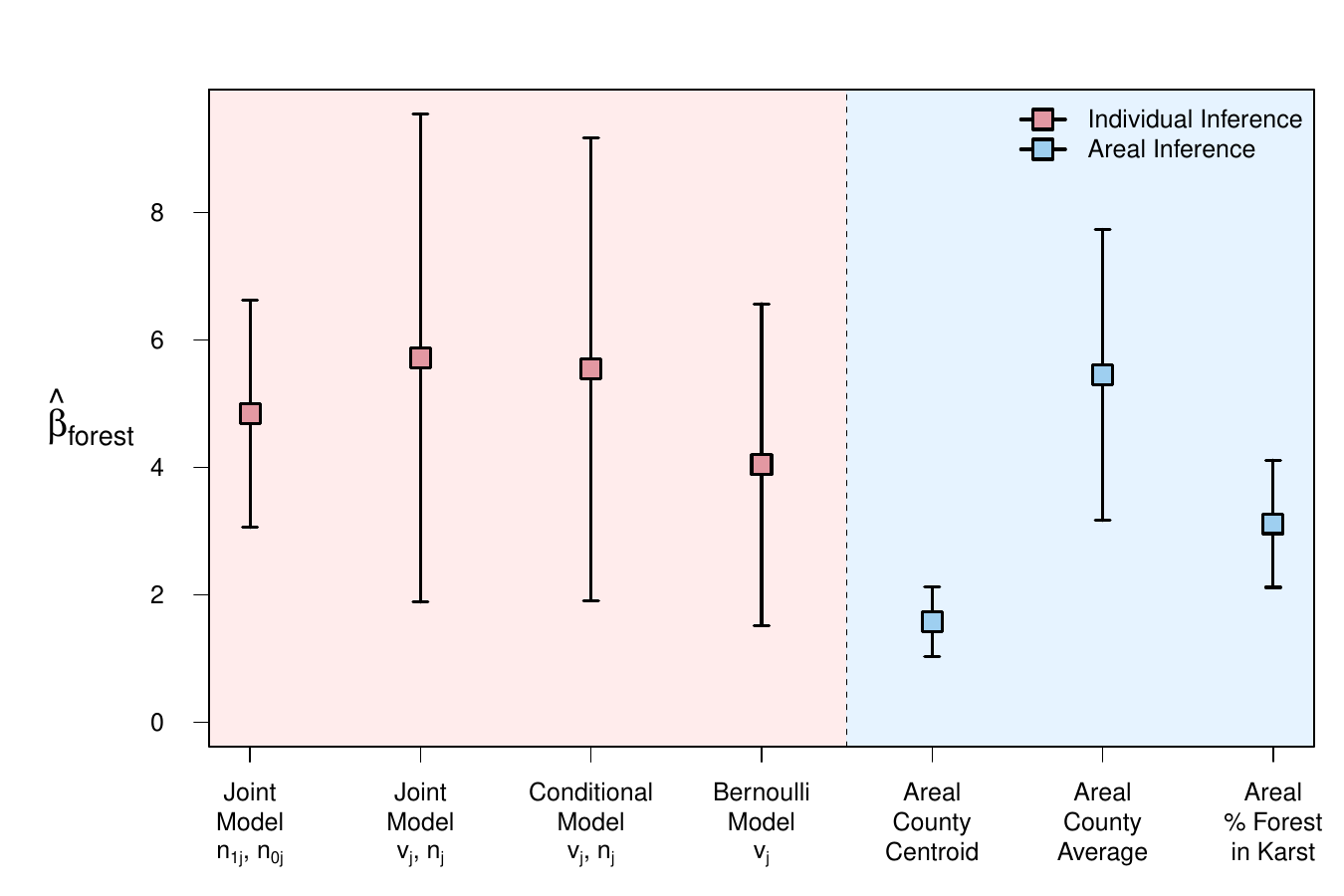}}
\caption{Binary regression model coefficient estimates and 95\% CIs for the spatial covariate `proportion of land classified as forest' (forest) that affects the probability of \textit{P. destructans} infection for cave-hibernating bats in the northeastern United States (see \textbf{Figure 1} for visual). Estimates were obtained from the joint model for $n_{1j}$ and $n_{0j}$ in (\ref{eq:COS_pos}-\ref{eq:COS_neg}),  the joint model for $v_{j}$ and $n_{j}$ in  (\ref{eq:COS_all}) and (\ref{eq:latent2.bivar.ipp}),  the conditional model for $v_{j}$ given $n_{j}$ in (\ref{eq:latent2.bivar.ipp}), and the Bernoulli model for $v_{j}$ in (\ref{eq:latent1.bivar.ipp}) that were fit using the respective data types. Here, $n_{1j}$ is the number of observations in the $j^{\mathrm{th}}$ county that tested positive or suspect positive for WNS, $n_{0j}$ is the number of observations in the $j^{\mathrm{th}}$ county that tested negative, $n_j$ is the total number of observations in the $j^{\mathrm{th}}$ county, and $v_j = \mathrm{I}(n_{1j}>0)$.  Also, using data that consists of the binary indicators ($v_{j}$), we give the areal-level results for logistic regression models that have the covariates of county centroid value of forest (Areal County Centroid), county averaged forest (Areal County Average), and county averaged forest in karst landscape (Areal $\%$ Forest in Karst). We delineate which models can recover individual-level inference (pink) and which are suited to areal-level inference (blue). For each model, we give the coefficient estimate (box) followed by the 95\% CI limits (whisker ends). 
}
\label{wns_analysis}
\end{figure}

Our results show that the proposed regression models give similar inference to each other regardless of the type of data or level of aggregation, as long as the appropriate model is used (see \textbf{Figure \ref{wns_analysis}} for comparisons and the web-based appendix for additional plots). The joint model for $n_{1j}$ and $n_{0j}$ from (\ref{eq:COS_pos}-\ref{eq:COS_neg}) provided the most precise estimates and matched the distribution of the available WNS data.
As a result, the joint model for $n_{1j}$ and $n_{0j}$ provides the most efficient individual-level inference among the proposed models. This is unsurprising because the data, which are Type C, contain the most information (see \textbf{Table 1}). 

The results for the logistic regression models fit to Type E data differed among themselves substantially, although the 95$\%$ CIs for $\hat{\beta}_{forest}$ overlapped between two pairs of the three models. While it would be tempting to compare the results from the logistic regression models fit to Type E data against the models that produce individual-level inference, it would be fallacious to do so (Piantadosi, Byar, and Green, 1988; Gotway and Young 2002). %This is because the areal-level data and models target county-level inference while our proposed models target individual-level inference.

%However, when true locations are unavailable because of differential privacy protection, estimates for individual level effect of proportion of land cover as forest would be impossible to obtain by ordinary means. 

\section{Discussion}
\label{s:discuss}

Our results demonstrated that models based on the proposed distributional results were capable of recovering individual-level inference on spatial covariates from aggregated binary data. As the degree of data aggregation increases, from Type C data to Type E, the relative efficiency of slope parameter estimates and intercept estimates decreases (see web-based appendix for additional results). Further, the probability of obtaining extreme values of coefficient estimates and standard errors from the proposed models increases as aggregation increases from Type A to Type E data. However, even without more specific information than an indicator variable (i.e., Type E data) for each county, our results show that it may be possible to recover individual-level inference.

In many situations, such as our WNS surveillance data, data curators will be unable to release the exact locations of binary data (i.e., Type A data). Likewise, there will be many situations where data curators may be unwilling or unable to release Type C aggregated data because the data contain too much specific information to adequately protect privacy. The next level of privacy protection that enables individual-level inference comes from releasing the number of observations in each subregion ($n_j$) and an indicator variable for each subregion ($v_j = \mathrm{I}(n_{1j}>0)$). Releasing $n_j$ and $v_j$ would provide the data required to fit models based on (\ref{eq:latent2.bivar.ipp}) and the joint density of (\ref{eq:COS_all}) and (\ref{eq:latent2.bivar.ipp}). We note that inference from the joint model for $v_j$ and $n_j$ is usually preferable  in practice if $n_{1j}$ and $n_{0j}$ are unavailable. This is because parameter estimates from the joint model for $v_j$ and $n_j$ are more efficient than that of the conditional model. The model for Type E data based on  (\ref{eq:latent1.bivar.ipp}) has an increased probability of providing extreme coefficient estimates and large or infinite standard errors for some situations (similar to complete separation in binary regression models). However, if auxiliary information is available about $\lambda(\cdot)$ (e.g. the sampling design for the study or a point estimate for $\lambda(\cdot)$), models based on (\ref{eq:latent1.bivar.ipp}) would have a higher probability of being useful (i.e., estimates may not be extreme and confidence intervals may be of reasonable width). 
%We caution against using the same spatial covariates in $\lambda(\mathbf{s})$ and $p(\mathbf{s})$ in models based on (\ref{eq:latent1.bivar.ipp}) because of lack of identifiability of the parameters. 
In general, if individual-level inference is required, we recommend that practitioners fit the appropriate model for the type of aggregated data that is available to them. If the standard errors are large for the parameters of interest in the appropriate model, we recommend applying standard techniques to address complete separation (e.g., a Firth correction; Firth, 1993). %Alternatively, we recommend that practitioners revert to traditional methods to make areal-level inference.

%Another application for the proposed transformed binary regression models is the area of spatial capture/recapture. As spatial capture/recapture data are becoming increasingly massive, a method to improve computation times is to spatially aggregate the data. Milleret et al. (2018) noted that spatial aggregation introduced bias into the analysis, akin to location error (see also Walker et al., 2020). The transformed binary regression models proposed in this paper are able to correct the bias introduced by aggregation and provide individual inference on spatial covariates.

Two issues linger from our disease risk factor analysis. First, in some disease risk factor analyses there may be a need to account for spatial correlation among the responses.  A spatial random effect may be added to the models proposed in this paper, either in the specification for $\lambda(\mathbf{s})$, or $p(\mathbf{s})$, or both (e.g., Diggle, Tawn, and Moyeed, 1998), as follows:

\begin{align}
\log(\lambda(\mathbf{s}))= &\alpha_0+\mathbf{z}(\mathbf{s})^\prime \boldsymbol{\alpha}+\eta(\mathbf{s}), \\
\mathrm{logit}(p(\mathbf{s}))= &\beta_0+\mathbf{x}(\mathbf{s})^\prime \boldsymbol{\beta}+\gamma(\mathbf{s}),
%\begin{bmatrix}
%\eta(\mathbf{s}_1)\\ 
%\vdots \\ 
%\eta(\mathbf{s}_n)\\ 
%\gamma(\mathbf{s}_1)\\ 
%\vdots \\ 
%\gamma(\mathbf{s}_n)
%\end{bmatrix}
%&\sim \mathrm{GP}(\mathbf{0}, \begin{bmatrix}
 %\boldsymbol{\Sigma}_\eta& \mathbf{0}\\ 
% \mathbf{0}& \boldsymbol{\Sigma}_\gamma
%\end{bmatrix}).
%\eta(\mathbf{s}) \sim & \mathrm{N}(0, \sigma^2_\eta), \\
%\gamma(\mathbf{s}) \sim & \mathrm{N}(0, \sigma^2_\gamma),
\end{align}

 \noindent where each value of $\eta(\mathbf{s})$ and $\gamma(\mathbf{s})$ is assumed to follow a multivariate normal distribution, as follows:
 
\begin{align}
\begin{bmatrix}
\eta(\mathbf{s}_1)\\ 
\vdots \\ 
\eta(\mathbf{s}_n)\\ 
\gamma(\mathbf{s}_1)\\ 
\vdots \\ 
\gamma(\mathbf{s}_n)
\end{bmatrix}
&\sim \mathrm{N}(\begin{bmatrix}\mathbf{0}\\ \mathbf{0} \end{bmatrix}, \begin{bmatrix}
 \boldsymbol{\Sigma}_\eta& \boldsymbol{\Sigma}_{\eta\gamma}\\ 
 \boldsymbol{\Sigma}_{\gamma\eta}& \boldsymbol{\Sigma}_\gamma
\end{bmatrix}).
\end{align}
 \noindent  Here, $\boldsymbol{\Sigma}_\eta$ and $\boldsymbol{\Sigma}_\gamma$ are block diagonal components of the covariance matrix and $\boldsymbol{\Sigma}_{\eta\gamma}=\boldsymbol{\Sigma}_{\gamma\eta}^\prime$ is an $n\times n$ block of zeros. Although practitioners could perform standard visual model checking procedures (e.g., semivariogram) to determine if spatial auto-correlation occurs in either the location data or the binary marks, we are unaware of how these techniques could be applied to aggregated data. Instead, we recommend that practitioners fit the proposed models with a spatial random effect(s), and then again without, and perform model selection (Burnham and Anderson, 2002). 

The second common issue for disease risk factor analyses is that collection of opportunistic disease surveillance data is often and likely the result of preferential sampling. Preferential sampling arises if $\eta(\mathbf{s})$ and$\gamma(\mathbf{s})$ from (12-14) are correlated, or when the off-diagonal blocks of the covariance matrix are non-zero. Including a spatial random effect is therefore a straightforward way to account for preferential sampling that may be present when using any of the models included in our paper (Diggle et al., 2010b). Adapting assumptions 1-3 from Diggle et al. (2010b) to our notation from (4) and assuming that (4) is specified with spatial random effects:

\begin{itemize}

\item[\textit{1.}] $\eta(\mathbf{s}) \sim \mathrm{N}(\mathbf{0}, \boldsymbol{\Sigma}_\eta)$, where $\eta(\mathbf{s})$ is a spatial random effect assumed to follow a multivariate normal distribution and $\mathbf{s}$ is the coordinate vector in the study area $\mathcal{S}$ (i.e., $\mathbf{s}\subseteq \mathcal{S}$).

\item[\textit{2.}] $\mathbf{U} \sim \mathrm{IPP}(\lambda(\mathbf{s}))$ where $\mathbf{U}\equiv (\mathbf{u}_1,\mathbf{u}_2,...,\mathbf{u}_n)^\prime$ is a matrix of locations for the tested bats generated from an inhomogeneous Poisson point process with $\log(\lambda(\mathbf{s}))=\alpha_0+\mathbf{z}(\mathbf{s})^\prime \boldsymbol{\alpha}+\theta \eta(\mathbf{s})$ and $\theta$ as a scaling parameter.

\item[\textit{3.}] $y_i \sim \mathrm{Bern}(p(\mathbf{u}_i)),$ where $y_i$ is the $i^{\mathrm{th}}$ observation, $\mathbf{u}_i$ is the location of the $i^{\mathrm{th}}$ bat, and $g(p(\mathbf{s}))=\beta_0+ \mathbf{x}(\mathbf{s})^\prime \boldsymbol{\beta} + \eta(\mathbf{s})$. For our purposes, $g(\cdot)$ is the logit link.
\end{itemize}

\noindent Following Diggle et al. (2010b), the model specified in items 1-3 accounts for preferential sampling.
 %We save the implementation of spatial random effects and the study of how they impact the proposed models as future work.

%Our simulation study and data example specifically use a logit link for the binary regression model. However, the COS technique is easily modified to incorporate other link functions (e.g., probit). Our approach could also accommodate other types of predictors, like kernel averaged predictors (Heaton and Gelfand, 2011), that account for the influence of surrounding regions. Additionally, one need not assume a linear relationship in the intensity function as shown in ($\ref{eq:loglink}$); a smooth function such as a semi-parametric or kernel density estimator may be used for the intensity function, $\lambda(\cdot)$ (Yue and Loh, 2011).  The proposed models can also be modified for data that are spatially and temporally aggregated. We refer to these options as modifications rather than extensions because these changes may be made without extending the basic framework we provide. 

Lastly, non-spatial individual-level covariates (e.g., sex or age) can be included in models for Type B and C data (e.g., Walker et al., 2020). However, due to the constraints inherent in the aggregation process for Type D and E data, it is not likely that non-spatial, individual-level covariates would be available. A future contribution might incorporate non-spatial, aggregated individual-level covariates (e.g., average age of tested individuals in a county) into the proposed transformed models for data Types D and E. Furthermore, Taylor, Andrade-Pacheco, and Sturrock (2018) and Heaton et al. (2020) showed it may be possible to relax the assumption of a discretized partition of the study area that normally applies to models that include a COS transformation. Relaxing this assumption would accommodate overlapping and uncertain subregion boundaries.

\section*{Acknowledgements}
We thank all state, federal and other partners for submitting samples and the USGS National Wildlife Health Center (Madison) for processing the samples. We thank the associate editor and two anonymous referees from \textit{Spatial Statistics} for their valuable feedback. We likewise thank Dr. Kathi Irvine for her comments via a Fundamental Science Practices (FSP) review. We acknowledge support for this research from USGS G18AC00317 and G16AC00413. Any use of trade, firm, or product names is for descriptive purposes only and does not imply endorsement by the U.S. Government. Declaration of Interest: None.

\section*{Supporting Information}
%The Web Appendix and Figures referenced in Section 3 are available with this paper at the Biometrics website on Wiley Online Library. 
The \texttt{.R} files referenced in Sections 3 and 4 are available within the \texttt{Recovering Inference.zip} from the Supplementary Material. The disease surveillance data used in this paper are available in the Supplementary Material and from the data release Ballmann et al. (2021). These data were provided by the U.S. Geological Survey, National Wildlife Health Center from a database that is continuously updated (accessed on Aug 22, 2019). Updated versions of the data may be requested from Anne Ballmann (aballmann@usgs.gov) with the permission of the National Wildlife Health Center and contributing partner agencies.

%\vspace*{-8pt}

%\section*{References}

\end{document}